\documentclass[a4paper,fleqn]{cas-dc}

\usepackage[numbers]{natbib}

\usepackage{algorithm} 
\usepackage{algpseudocode}
\algrenewcommand\algorithmicindent{1.2em}

\def\tsc#1{\csdef{#1}{\textsc{\lowercase{#1}}\xspace}}
\tsc{WGM}
\tsc{QE}
\tsc{EP}
\tsc{PMS}
\tsc{BEC}
\tsc{DE}

\makeatletter
\def\printorcid{}
\makeatother

\begin{document}
\let\WriteBookmarks\relax
\def\floatpagepagefraction{1}
\def\textpagefraction{.001}

\shorttitle{Safe Control and Learning Using Generalized Action Governor}

\shortauthors{Peiyuan Fang et~al.}

\title [mode = title]{Safe Control and Learning Using Generalized Action Governor}                      



%

\author[1]{Peiyuan Fang}[bioid=1]
\fnmark[1]
\credit{Writing – review \& editing, Writing – original draft, Visualization, Validation, Software, Resources, Project administration, Methodology}

\author[1]{Weiqi Zhang}[bioid=2]
\fnmark[1]
\credit{Writing – review \& editing, Writing – original draft, Visualization, Validation, Software, Resources, Project administration, Methodology}

\author[1]{Lu Xiong}[bioid=3]
\credit{Visualization, Funding acquisition, Data curation, Conceptualization}

\author[1]{Nan Li}[bioid=4]
\cormark[1]
\credit{Writing – review \& editing, Writing – original draft, Visualization, Validation, 
Resources, Formal analysis, Data curation, Software}

\author[1]{Yanjun Huang}[bioid=5]
\credit{Writing – review \& editing, Supervision, Conceptualization}

\author[2]{Yutong Li}[bioid=6]
\credit{Supervision, Resources, Formal analysis}

\author[2]{Ilya Kolmanovsky}[bioid = 7]
\credit{Supervision, Methodology}

\author[3]{Anouck Girard}[bioid = 8]
\credit{Supervision, Methodology}

\author[4]{H. Eric Tseng}[bioid = 9]
\credit{Supervision, Resources, Software}

\author[5]{Dimitar Filev}[bioid = 10]
\credit{Supervision, Resources, Software}

\affiliation[1]{organization={School of Automotive Studies, Tongji University},
    city={Shanghai},
    postcode={201804}, 
    country={China}}

\affiliation[2]{organization={Department of Aerospace Engineering, University of Michigan},
    city={Ann Arbor},
    postcode={48109}, 
    state={MI},
    country={USA}}

\affiliation[3]{organization={Department of Aerospace Engineering, Embry-Riddle Aeronautical University},
    city={Daytona Beach},
    postcode={32114}, 
    state={FL},
    country={USA}}

\affiliation[4]{organization={Department of Electrical Engineering, University of Texas at Arlington},
    city={Arlington},
    postcode={76019}, 
    state={TX},
    country={USA}}
    
\affiliation[5]{organization={Hagler Institute for Advanced Study, Texas A\&M University},
    city={College Station},
    postcode={77840}, 
    state={TX},
    country={USA}}

\cortext[cor1]{Corresponding author: N. Li (li\_nan@tongji.edu.cn)}

\fntext[fn1]{These authors contributed equally.}

\begin{abstract}
This paper introduces the Generalized Action Governor (AG), a supervisory scheme that augments a nominal closed-loop system with the capability to enforce state and input constraints through online action adjustment. We develop a generalized AG theory for discrete-time systems under bounded uncertainties, and relax the usual requirement of positive invariance to returnability of a safe set. Based on the theory, we present tailored AG design procedures for linear systems and for discrete systems with finite state and action spaces. We further study safe online learning enabled by the AG and present two safe learning strategies, namely safe $Q$-learning and safe data-driven Koopman operator-based control, both integrated with the AG to guarantee constraint satisfaction during learning. Numerical results illustrate the proposed methods.
\end{abstract}




\maketitle

\section{Introduction}\label{sec:intro}

Safety is a major concern in the development and operation of autonomous systems. Many safety specifications can be expressed as constraints on the system state and control variables. Control methods that can explicitly handle constraints include model predictive control (MPC) \cite{camacho2013model,borrelli2017predictive}, reachability-based methods \cite{ahn2020reachability,herbert2021scalable}, control Lyapunov/barrier~functions \cite{tee2009barrier,ames2016control}, reference/action governors \cite{garone2017reference,kolmanovsky2022protecting}, and others.

Inspired by the general idea of reference governors (RGs), the AG is a recently proposed scheme that enhances a nominal closed-loop system with the capability of strictly handling constraints \cite{li2020action,li2021robust}. Unlike the RG which is placed in front of a nominal controller and supervises the reference inputs to the closed-loop system \cite{garone2017reference,kolmanovsky2022protecting}, the AG is placed after the controller and supervises the control actions generated by the controller, adjusting unsafe actions to safe ones. An advantage of placing a supervisory scheme after the controller is that it allows for the nominal controller to be modified, such as through online learning, without requiring a redesign of the supervisory scheme \cite{li2021safereinforcementlearningusing}. The AGs in \cite{li2020action,li2021robust,li2021safereinforcementlearningusing} are designed based on discrete-time models and methods that exploit set-based computations. This distinguishes them from control Lyapunov/barrier functions, which are most frequently based on continuous-time models \cite{tee2009barrier,ames2016control} (although discrete-time formulations do exist \cite{zeng2021safety,xiong2022discrete}) and typically do not use set-based methods. The AG was first developed for discrete-time linear models in \cite{li2020action} and then extended to uncertain piecewise-affine models in~\cite{li2021robust}. 

The first part of this paper aims to generalize the AG theory. Instead of focusing on specific system models, such as linear or piecewise-affine models as in \cite{li2020action,li2021robust,li2021safereinforcementlearningusing}, we make minimal assumptions about the system and describe both the offline design and online operation procedures of the AG in this more general framework. Next, we introduce tailored design procedures and algorithms for linear systems, as well as for discrete systems with finite state and action spaces, highlighting the unique characteristics of each. Note that the procedure and algorithm for linear systems introduced in this paper are different from those of \cite{li2020action} -- the ones introduced in this paper are derived based on the generalized AG theory developed in this paper, which are computationally efficient and scalable but restricted to convex constraints, while those of \cite{li2020action} are not. Moreover, we introduce a novel procedure and algorithm that handle discrete systems.

As many autonomous systems operate under uncertain or changing conditions (e.g., due to environmental factors or component aging), it is highly desirable for these systems to have online learning capabilities that adapt control parameters based on real-time data. Such learning must be performed safely, meaning that constraints must be satisfied throughout the learning process to ensure the system's stability and safety. However, most conventional learning approaches, including most reinforcement learning (RL) algorithms \cite{sutton2018reinforcement}, do not have the ability to strictly respect constraints during learning. As a matter of fact, this has been a major impediment to using these approaches for online learning. To overcome this obstacle, a learning algorithm may be integrated with a constraint handling scheme to realize safe learning. For instance, safe RL using MPC is proposed in \cite{zanon2020safe,li2020robust,wabersich2021predictive} and using control Lyapunov/barrier functions in \cite{cheng2019end,choi2020reinforcement,marvi2021safe}. Integrating $Q$-learning with an AG to realize safe $Q$-learning was also proposed in our previous conference paper \cite{li2021safereinforcementlearningusing}, which highlights the importance of ensuring safety during the learning process. These approaches are based on a similar idea of projecting the RL action on a safe set, as described in \cite{gros2020safe}. In the second part of this paper, we extend the discussion on safe online learning using the AG. To make the paper self-contained, we first re-elaborate the integration of $Q$-learning and the AG for safe $Q$-learning, where we provide additional details beyond our conference paper \cite{li2021safereinforcementlearningusing}. Then, we introduce a new safe learning strategy based on data-driven Koopman control. Koopman operator-based control, where control is determined based on a data-driven Koopman linear model of a nonlinear system, is an emerging area of research \cite{korda2018linear,bruder2020data}. In this paper, we propose to integrate it with the AG as an alternative safe learning strategy to safe RL. To the best of our knowledge, such a safe learning strategy based on data-driven Koopman control has not been proposed before. 

This paper develops a generalized AG theory that is independent of the specific form of the system model, enabling its application to a broader class of systems. Unlike previous AG design approaches that are limited to linear or piecewise-affine models, the proposed theory relaxes the requirement for the positive invariance of a safe set, allowing for more flexible design. This relaxation leads to several important consequences: i) it enables the design of a returnable safe set for systems where positively invariant safe sets are difficult to design \cite{gilbert2002nonlinear,kolmanovsky2022protecting}; ii) it reduces the complexity of safe set representations, which can be advantageous for both memory storage and online computation \cite{kolmanovsky2022protecting}; and iii) for discrete systems, it allows for the design of a safe set using the algorithm proposed in Section~\ref{sec:GAG_discrete}. Based on the generalized AG theory, we present tailored AG design approaches for linear systems and discrete systems with bounded uncertainties. For linear systems, we show that the maximum output admissible set (MOAS), which has been studied in the context of reference governors (RGs) \cite{garone2017reference,kolmanovsky2022protecting}, can be used to define the safe set for AG to handle constraints, leading to a new, computationally efficient AG design approach for linear systems with convex constraints. For discrete systems with finite state and action spaces, we propose a novel algorithm for computing the safe set. Furthermore, we apply the generalized AG to safe online learning by integrating the AG with a new learning strategy based on Koopman operator-based control. This approach extends the use of AG in safe reinforcement learning (RL) and introduces a new method for improving control performance using real-time data for uncertain systems.

The contributions of this paper are:

1) A generalized AG theory applicable to more general systems, allowing for more flexible safe set design.

2) Tailored AG design methods for linear and discrete systems with bounded uncertainties, improving computational efficiency.

3) A novel safe learning strategy integrating Koopman operator-based control with AG for safe online learning.

The organization of this paper is as follows: In Section~\ref{sec:MA}, we introduce the basic models and assumptions for the AG design. In Section~\ref{sec:GAG}, we present the generalized AG, analyze its properties, and introduce tailored approaches for linear and discrete systems. In Section~\ref{sec:SOL}, we discuss the application of the generalized AG to safe online learning. We then use an example to illustrate the previous developments in Section~\ref{sec:Ex} and conclude the paper in Section~\ref{sec:conclude}.

\section{Models and Assumptions}\label{sec:MA}

This paper considers systems whose dynamics can be represented by the following discrete-time model:
\begin{equation}\label{equ:system}
x(t+1) = f\big(x(t), u(t), w(t)\big)
\end{equation}
where $t \in \mathbb{Z}_{\ge 0}$ denotes the discrete time; $x(t) \in \mathcal{X}$ represents the system state at time $t$, taking values in the state space $\mathcal{X}$; $u(t) \in \mathcal{U}$ represents the control action at time $t$, taking values in the action space $\mathcal{U}$; $w(t) \in \mathcal{W}$ represents an uncertainty such as an unmeasured external disturbance, which is assumed to take values in a known bounded set $\mathcal{W}$; and $f: \mathcal{X} \times \mathcal{U} \times \mathcal{W} \to \mathcal{X}$ is the state transition function. At this stage, we make no further assumptions on the spaces $\mathcal{X}$, $\mathcal{U}$, and $\mathcal{W}$ and on the function~$f$. For instance, $\mathcal{X}$, $\mathcal{U}$, and $\mathcal{W}$ can be continuous or discrete spaces, and $f$ can be nonlinear.

This paper deals with safety-critical applications where it is assumed that the system operation must satisfy the following constraints on state and control variables:
\begin{equation}\label{equ:constraint}
    \big(x(t), u(t)\big) \in \mathcal{C}, \quad \forall t \in \mathbb{Z}_{\ge 0}
\end{equation}
where $\mathcal{C}$ is a subset of $\mathcal{X} \times \mathcal{U}$. A constraint on the state $x(t)$ can represent a process variable bound or collision avoidance; a constraint on the control action $u(t)$ can present an actuator power or range limit, which may be state-dependent.

In order to develop the safety supervisor termed {\it Generalized Action Governor (AG)} for enforcing \eqref{equ:constraint}, we assume there is a nominal control policy, $\pi_0: \mathcal{X} \times \mathcal{V} \to \mathcal{U}$, that is,
\begin{equation}\label{equ:policy0}
    u_0(t) = \pi_0 \big(x(t), v(t)\big)
\end{equation}
where $v(t) \in \mathcal{V}$ represents a reference signal, which typically corresponds to the current control objective (e.g., a setpoint for tracking) and is passed to the policy from a higher-level planner or a human operator. 

We write the closed-loop system under the nominal policy $\pi_0$ as
\begin{equation}\label{equ:system0}
    x(t+1) = f_{\pi_0}\big(x(t), v(t), w(t)\big) = f\big(x(t), \pi_0(x(t), v(t)), w(t)\big).
\end{equation}
For a given initial condition $x(0)=x$, a constant reference $v(\tau)\equiv v$, and a disturbance sequence $w(\cdot)=\{w(\tau)\}_{\tau=0}^{\infty}$ with $w(\tau)\in\mathcal{W}$ for all $\tau\in\mathbb{Z}_{\ge 0}$, we denote the resulting state at time $t$ by $\phi_{\pi_0}(t\,|\,x,v,w(\cdot))$.
This nominal closed-loop system is assumed to exhibit well-behaved responses under constant references so that a nonempty safe and returnable set (defined in \eqref{equ:safety} and \eqref{equ:returnability}) exists. For instance, under a constant $v(t)\equiv v\in\mathcal{V}$, the state $x(t)$ may converge to a steady state corresponding to $v$, denoted by $x_v(v)$, i.e., $x(t)\approx x_v(v)$ as $t\to\infty$. Note that we do not require the nominal policy $\pi_0$ to satisfy the safety constraints in \eqref{equ:constraint} for all $v\in\mathcal{V}$, nor do we assume it is an optimal policy. Therefore, for many systems such a policy is easier to design than one that simultaneously achieves stability and constraint handling. For instance, for linear systems $\pi_0$ may be designed based on PID or linear quadratic regulator (LQR).

We now consider a nonempty compact set $\Pi_{\pi_0}\subseteq \mathcal{X}\times\mathcal{V}$ of pairs $(x,v)$ that satisfies the following two properties:

{\it A (Safety).} For all $(x,v) \in \Pi_{\pi_0}$ and any disturbance sequence $w(\cdot)=\{w(\tau)\}_{\tau=0}^{\infty}$ with $w(\tau)\in\mathcal{W}$ for all $\tau\in\mathbb{Z}_{\ge 0}$,
\begin{equation}\label{equ:safety}
\big(\phi_{\pi_0}(t \,|\, x,v,w(\cdot)), \pi_0 \big(\phi_{\pi_0}(t \,|\, x,v,w(\cdot)), v\big) \big) \in \mathcal{C}
\end{equation}
for all $t \in \mathbb{Z}_{\ge 0}$.

{\it B (Returnability).} For all $(x,v) \in \Pi_{\pi_0}$ and any disturbance sequence $w(\cdot)=\{w(\tau)\}_{\tau=0}^{\infty}$ with $w(\tau)\in\mathcal{W}$ for all $\tau\in\mathbb{Z}_{\ge 0}$, there exists $\hat{t}=\hat{t}(x,v,w(\cdot))\in\mathbb{Z}_{\ge 1}$ such that
\begin{equation}\label{equ:returnability}
\big(\phi_{\pi_0}(\hat{t} \,|\, x,v,w(\cdot)), v\big)  \in \Pi_{\pi_0}.
\end{equation}

The returnability of $\Pi_{\pi_0}$ means that state trajectories of the nominal closed-loop system \eqref{equ:system0} for constant $v$ beginning in $\Pi_{\pi_0}$ eventually return to $\Pi_{\pi_0}$. If for all $(x,v) \in \Pi_{\pi_0}$ and all disturbance sequences $w(\cdot)$ taking values in $\mathcal{W}$ the return time in \eqref{equ:returnability} can be chosen as $\hat{t}=1$, then $\Pi_{\pi_0}$ is {\it positively invariant}. Thus, positively invariant sets belong to the larger class of returnable sets. Given a nominal control policy $\pi_0$, a variety of tools exist for computing or approximating a set $\Pi_{\pi_0}$ that satisfies the two properties -- {\it safety} and {\it returnability} (or the stronger property {\it positive invariance}). In Section~\ref{sec:GAG}, we introduce two examples, one for linear systems and the other for discrete nonlinear systems.

\section{Generalized Action Governor}\label{sec:GAG}

The generalized AG enforces \eqref{equ:constraint} by adjusting actions according to the following algorithm:
\begin{equation}\label{equ:AG_1}
    u(t) = \begin{cases} 
    \hat{u}(t), & \text{if \eqref{equ:AG_2} is feasible}, \\
    \pi_0\big(x(t), \hat{v}(t)\big), & \text{otherwise},
    \end{cases}
\end{equation}
where $\hat{u}(t)$ is obtained by solving
\begin{subequations}\label{equ:AG_2}
\begin{align}
    \hat{u}(t) \in \text{argmin}_{u \in \mathcal{U}}\,\, & \text{\sf dist}_{x(t)}\big(u_1(t), u\big) \label{equ:AG_21} \\
    \text{s.t. } (x(t), u) & \in \mathcal{C} \label{equ:AG_22} \\
    f\big(x(t), u, w\big) & \in \text{proj}_x \big(\Pi_{\pi_0}\big),\, \forall w \in \mathcal{W}. \label{equ:AG_23}
\end{align}
\end{subequations}
The reference $\hat{v}(t)$ is selected as follows. If $x(t) \in \text{proj}_x \big(\Pi_{\pi_0}\big)$, then
\begin{subequations}\label{equ:AG_3}
\begin{align}
    \hat{v}(t) \in \text{argmin}_{v \in \mathcal{V}}\,\, & \text{\sf dist}_{x(t)}\big(u_1(t), \pi_0(x(t), v) \big) \label{equ:AG_31} \\
    \text{s.t. } (x(t), v) & \in \Pi_{\pi_0}. \label{equ:AG_32}
\end{align}
\end{subequations}
Otherwise, we set $\hat{v}(t)=\hat{v}(t-1)$.

In \eqref{equ:AG_2} and \eqref{equ:AG_3}, $u_1(t)$ denotes the action before adjustment. It does not have to be generated by the nominal policy $\pi_0$; it may come from another policy $\pi_1$ currently in use, or from an exploration action used to update $\pi_1$ through learning. The function $\text{\sf dist}_x(\cdot,\cdot)$ measures the deviation between $u_1(t)$ and the adjusted action $u$ (or $\pi_0(x(t),v)$), thereby minimizing the modification introduced by the supervisor. A typical choice is $\text{\sf dist}_x(u_1,u)=\|u_1-u\|$, where $\|\cdot\|$ denotes a norm; the subscript $x$ indicates that the distance metric may be state-dependent. The operator $\text{proj}_x$ projects the set $\Pi_{\pi_0}$ onto the state space $\mathcal{X}$.

The key idea of \eqref{equ:AG_1}--\eqref{equ:AG_3} is as follows. At each $t \in \mathbb{Z}_{\ge 0}$, if there exists an action $u$ that satisfies the instantaneous constraint \eqref{equ:AG_22} and, for all $w\in\mathcal{W}$, steers the next state into $\text{proj}_x(\Pi_{\pi_0})$ as required by \eqref{equ:AG_23}, then the AG selects (among all such actions) one that is closest to $u_1(t)$ in the sense of \eqref{equ:AG_21}. If no such action exists, the AG switches to the backup nominal controller and applies $u(t)=\pi_0\big(x(t),\hat{v}(t)\big)$. In this case, if $x(t)\in \text{proj}_x(\Pi_{\pi_0})$ (equivalently, \eqref{equ:AG_3} is feasible by construction), $\hat{v}(t)$ is chosen to minimize the adjustment via \eqref{equ:AG_31} subject to \eqref{equ:AG_32}; otherwise, we keep $\hat{v}(t)=\hat{v}(t-1)$.

The generalized AG algorithm has the following properties:

{\it Proposition~1 (All-Time Safety).} If \eqref{equ:AG_2} is feasible at the initial time $t=0$, then the trajectory $(x(t),u(t))$ under AG supervision satisfies \eqref{equ:constraint} for all $t\in\mathbb{Z}_{\ge 0}$.

{\it Proof.} Let $\tau\in\mathbb{Z}_{\ge 0}$ be arbitrary. Since \eqref{equ:AG_2} is feasible at $t=0$, the set $\{t\in\{0,1,\ldots,\tau\}:\eqref{equ:AG_2}\ \text{is feasible at }t\}$ is nonempty; let
\[
\tau' := \max\{t\in\{0,1,\ldots,\tau\}:\eqref{equ:AG_2}\ \text{is feasible at }t\}.
\]
By \eqref{equ:AG_1} and \eqref{equ:AG_22}, we have $(x(\tau'),u(\tau'))=(x(\tau'),\hat{u}(\tau'))\in\mathcal{C}$. If $\tau'=\tau$, we are done. If $\tau'<\tau$, then by \eqref{equ:AG_1} and \eqref{equ:AG_23},
\[
x(\tau'+1)=f\big(x(\tau'),\hat{u}(\tau'),w(\tau')\big)\in \text{proj}_x(\Pi_{\pi_0}),
\]
which implies that \eqref{equ:AG_3} is feasible at $\tau'+1$. Similarly, let
\[
\tau'' := \max\{t\in\{\tau'+1,\ldots,\tau\}:\eqref{equ:AG_3}\ \text{is feasible at }t\}.
\]
By \eqref{equ:AG_32}, $(x(\tau''),\hat{v}(\tau''))\in\Pi_{\pi_0}$. Moreover, by the definitions of $\tau'$ and $\tau''$, \eqref{equ:AG_2} is infeasible over $[\tau'+1,\tau]$ and \eqref{equ:AG_3} is infeasible over $[\tau''+1,\tau]$. Hence, by \eqref{equ:AG_1} and the update rule of $\hat{v}(t)$, for all $t\in[\tau'',\tau]$ we have
\[
u(t)=\pi_0\big(x(t),\hat{v}(t)\big),\quad \hat{v}(t)=\hat{v}(\tau'').
\]
Consequently, $x(\tau)=\phi_{\pi_0}(\tau-\tau''\,|\,x(\tau''),\hat{v}(\tau''),w(\cdot))$ and $u(\tau)=\pi_0\big(x(\tau),\hat{v}(\tau'')\big)$. Using $(x(\tau''),\hat{v}(\tau''))\in\Pi_{\pi_0}$ and the safety property \eqref{equ:safety}, we obtain $(x(\tau),u(\tau))\in\mathcal{C}$. Since $\tau$ is arbitrary, \eqref{equ:constraint} holds for all $t\in\mathbb{Z}_{\ge 0}$. $\blacksquare$

{\it Proposition~2 (Eventual Feasibility).} If \eqref{equ:AG_2} or \eqref{equ:AG_3} is feasible at time $t=\tau$, then there exists a future time $\tau'>\tau$ such that \eqref{equ:AG_2} or \eqref{equ:AG_3} is feasible at $t=\tau'$.

{\it Proof.} If \eqref{equ:AG_2} is feasible at $t=\tau$, then by \eqref{equ:AG_1} and \eqref{equ:AG_23} we have
\[
x(\tau+1)=f\big(x(\tau),\hat{u}(\tau),w(\tau)\big)\in \text{proj}_x(\Pi_{\pi_0}),
\]
and hence \eqref{equ:AG_3} is feasible at $\tau'=\tau+1$.

Next, consider the case where \eqref{equ:AG_2} is infeasible while \eqref{equ:AG_3} is feasible at $t=\tau$. Then, by \eqref{equ:AG_32}, $(x(\tau),\hat{v}(\tau))\in\Pi_{\pi_0}$. Suppose, for contradiction, that neither \eqref{equ:AG_2} nor \eqref{equ:AG_3} is feasible for all future times $t>\tau$. Then by \eqref{equ:AG_1} and the update rule of $\hat{v}(t)$, for all $t\ge \tau$ we have $u(t)=\pi_0\big(x(t),\hat{v}(t)\big)$ and $\hat{v}(t)=\hat{v}(\tau)$. Consequently,
\[
x(t)=\phi_{\pi_0}(t-\tau\,|\,x(\tau),\hat{v}(\tau),w(\cdot)),\quad
u(t)=\pi_0\big(x(t),\hat{v}(\tau)\big).
\]
However, since $(x(\tau),\hat{v}(\tau))\in\Pi_{\pi_0}$ and $\Pi_{\pi_0}$ is returnable by \eqref{equ:returnability}, there exists a time $\hat{t}\ge \tau+1$ such that
\[
\big(x(\hat{t}),\hat{v}(\tau)\big)
=
\Big(\phi_{\pi_0}(\hat{t}-\tau\,|\,x(\tau),\hat{v}(\tau),w(\cdot)),\,\hat{v}(\tau)\Big)
\in \Pi_{\pi_0}.
\]
Therefore, \eqref{equ:AG_3} is feasible at $t=\hat{t}$, contradicting the assumed infeasibility for all $t>\tau$. Hence, there must exist a future time $\tau'>\tau$ such that \eqref{equ:AG_2} or \eqref{equ:AG_3} is feasible at $t=\tau'$. $\blacksquare$

It can be seen from the proof that the returnability of $\Pi_{\pi_0}$ plays a key role in establishing the eventual-feasibility result of Proposition~2.
While the AG algorithm \eqref{equ:AG_1}--\eqref{equ:AG_3} and Propositions~1--2 apply to general systems, we next present two AG design approaches tailored for linear systems and discrete systems, respectively.

\vspace{-2mm}
\subsection{Generalized action governor for linear systems}\label{sec:GAG_linear}

Suppose that \eqref{equ:system} is linear with additive uncertainty, i.e.,
\begin{equation}\label{equ:system_linear}
    x(t+1)=f\big(x(t),u(t),w(t)\big)=Ax(t)+Bu(t)+Ew(t),
\end{equation}
where $A$, $B$, and $E$ are matrices of appropriate dimensions, and $w(t)$ takes values in a compact polyhedral set $\mathcal{W}=\{w: Gw\le g\}$. We assume that the constraints in \eqref{equ:constraint} can be represented by linear inequalities through an output map
\begin{equation}\label{equ:constraint_linear_def}
    y(t)=Cx(t)+Du(t), \qquad y(t)\in\mathcal{Y}=\{y: Hy\le h\},
\end{equation}
equivalently defining
\begin{equation}\label{equ:constraint_linear}
    \mathcal{C}:=\{(x,u)\in\mathcal{X}\times\mathcal{U}: Cx+Du\in\mathcal{Y}\}.
\end{equation}

In this case, we consider a linear nominal policy of the form \eqref{equ:policy0},
\begin{equation}\label{equ:policy0_linear}
    u_0(t)=\pi_0\big(x(t),v(t)\big)=Kx(t)+Lv(t),
\end{equation}
which yields the closed-loop system
\begin{align}\label{equ:system0_linear}
    x(t+1) &= f_{\pi_0}\big(x(t),v(t),w(t)\big)=\tilde{A}x(t)+\tilde{B}v(t)+Ew(t), \nonumber\\
    y(t) &= \tilde{C}x(t)+\tilde{D}v(t),
\end{align}
where $\tilde{A}=A+BK$, $\tilde{B}=BL$, $\tilde{C}=C+DK$, and $\tilde{D}=DL$. The nominal policy \eqref{equ:policy0_linear} can be designed flexibly, but it is required that $\tilde{A}$ is Schur (i.e., all eigenvalues lie strictly inside the unit disk). As will be shown below, this policy is used to compute, offline, a set $\Pi_{\pi_0}$ with the desired safety and returnability properties; it is not applied as the online control input. In particular, we define

\begin{equation}\label{equ:O_inf_1}
\Pi_{\pi_0}=\tilde{\mathcal{O}}_{\infty}=\left(\bigcap_{t=0}^{t'}\mathcal{O}_t\right)\cap (\mathcal{X}\times\Omega'),
\end{equation}
where
\begin{align}\label{equ:O_inf_2}
\mathcal{O}_t
&=\Big\{(x,v): \tilde{C}\tilde{A}^t x+\tilde{F}_t v \in \mathcal{Y}_t\Big\}, \nonumber\\
\mathcal{Y}_t
&=\mathcal{Y}\sim \tilde{C}\left(\bigoplus_{k=0}^{t-1}\tilde{A}^kE\,\mathcal{W}\right), \\
\Omega'
&=\Big\{v: \tilde{F}v\in (1-\varepsilon)\mathcal{Y}_{t'}\Big\}. \nonumber\\
\intertext{and $\tilde{F}_t$ and $\tilde{F}$ are defined as}
\tilde{F}_t
&:= \tilde{C}(I-\tilde{A}^t)(I-\tilde{A})^{-1}\tilde{B}+\tilde{D}, \\
\tilde{F}
&:= \tilde{C}(I-\tilde{A})^{-1}\tilde{B}+\tilde{D}. 
\end{align}
with $0<\varepsilon\ll 1$. In \eqref{equ:O_inf_2}, $\oplus$ denotes the Minkowski sum and $\sim$ denotes the Pontryagin difference \cite{kolmanovsky1998theory}.
Under mild assumptions, there exists a finite $t'$ such that the set $\tilde{\mathcal{O}}_{\infty}$ in \eqref{equ:O_inf_1} (the maximum output admissible set, MOAS) is well-defined and is safe and positively invariant (hence returnable) \cite{kolmanovsky1998theory}, i.e.,
\begin{align}\label{equ:O_inf_3}
& (x(t),v(t))\in \tilde{\mathcal{O}}_{\infty}\ \Longrightarrow\\
& \big(\tilde{A}x(t)+\tilde{B}v(t)+Ew,\ v(t)\big)\in \tilde{\mathcal{O}}_{\infty},\ \forall w\in\mathcal{W}. \nonumber
\end{align}

Under the positive invariance of $\Pi_{\pi_0}=\tilde{\mathcal{O}}_{\infty}$, the generalized AG \eqref{equ:AG_1}--\eqref{equ:AG_3} simplifies to the following one-step optimization:
\begin{subequations}\label{equ:AG_4}
\begin{align}
    u(t)\in \text{argmin}_{u\in\mathcal{U}}\,\, & \text{\sf dist}_{x(t)}\big(u_1(t),u\big) \\
    \text{s.t. } Cx(t)+Du &\in \mathcal{Y}, \label{equ:AG_41}\\
    Ax(t)+Bu &\in \text{proj}_x\big(\tilde{\mathcal{O}}_{\infty}\big)\sim E\mathcal{W}. \label{equ:AG_42}
\end{align}
\end{subequations}
This simplification follows from the next result.

{\it Proposition~3 (Recursive Feasibility).} Under the positive invariance of $\tilde{\mathcal{O}}_{\infty}$ in \eqref{equ:O_inf_3}, if \eqref{equ:AG_4} is feasible at time $t$, then it is feasible at time $t+1$.

{\it Proof.} If \eqref{equ:AG_4} is feasible at time $t$, then by \eqref{equ:AG_42} we have $Ax(t)+Bu(t)\in \text{proj}_x(\tilde{\mathcal{O}}_{\infty})\sim E\mathcal{W}$, which implies $x(t+1)=Ax(t)+Bu(t)+Ew(t)\in \text{proj}_x(\tilde{\mathcal{O}}_{\infty})$. Hence, there exists $v$ such that $(x(t+1),v)\in \tilde{\mathcal{O}}_{\infty}$. Consider $u=\pi_0(x(t+1),v)=Kx(t+1)+Lv$. Since $(x(t+1),v)\in \tilde{\mathcal{O}}_{\infty}\subseteq \mathcal{O}_0$, the definition of $\mathcal{O}_0$ in \eqref{equ:O_inf_2} yields $\tilde{C}x(t+1)+\tilde{D}v\in\mathcal{Y}$, i.e., $Cx(t+1)+Du\in\mathcal{Y}$, so \eqref{equ:AG_41} holds at $t+1$. Moreover, positive invariance \eqref{equ:O_inf_3} implies that $(\tilde{A}x(t+1)+\tilde{B}v+Ew,\ v)\in \tilde{\mathcal{O}}_{\infty}$ for all $w\in\mathcal{W}$; equivalently, $A x(t+1)+Bu+Ew\in \text{proj}_x(\tilde{\mathcal{O}}_{\infty})$ for all $w\in\mathcal{W}$, which is the same as $Ax(t+1)+Bu\in \text{proj}_x(\tilde{\mathcal{O}}_{\infty})\sim E\mathcal{W}$. Thus \eqref{equ:AG_42} holds at $t+1$ as well, and $u$ is a feasible solution of \eqref{equ:AG_4} at $t+1$. $\blacksquare$

Note that the computation of $\tilde{\mathcal{O}}_{\infty}$ for linear systems is simple and scalable \cite{kolmanovsky1998theory}. Therefore, the approach in this subsection provides a computationally efficient and scalable method for designing AGs for linear systems, and it differs from the approach of \cite{li2020action}.

\subsection{Generalized action governor for discrete systems}\label{sec:GAG_discrete}

We next consider the case where the spaces $\mathcal{X}$, $\mathcal{U}$, $\mathcal{W}$, and $\mathcal{V}$ are all finite, and the transition function $f$ maps between finite spaces. In this setting, \eqref{equ:system} is referred to as a \emph{discrete system}. This case is of interest for two main reasons. First, when the uncertainty $w(t)$ is modeled as a random variable with a probability distribution over $\mathcal{W}$, the dynamics in \eqref{equ:system} induce a Markov decision process (MDP) representation with transition probabilities determined by the distribution of $w(t)$ \cite{li2019stochastic}. MDPs are fundamental models for sequential decision-making and are widely used in many disciplines, including robotics and manufacturing \cite{thrun2005probabilistic}; they also admit convenient graph representations. Second, through appropriate discretization, a general nonlinear system can be approximated by a discrete system on finite spaces, and many control synthesis techniques (e.g., dynamic programming) are based on such discrete approximations \cite{puterman2014markov}. Therefore, the techniques presented in this subsection can also be used to treat general nonlinear systems in an approximate manner.

\begin{algorithm}
	\caption{Computation of $\Pi_{\pi_0}$}
	\begin{algorithmic}[1]
	    \State Initialize $\mathcal{D}_{xv}^+ \gets \Pi_{\pi_0}^0$, $\mathcal{D}_{xv}^- \gets \emptyset$, $\mathcal{D}_{xv}^{\text{remain}} \gets (\mathcal{X}\times\mathcal{V})\setminus \Pi_{\pi_0}^0$, and $k \gets 0$;
	    \While{$\mathcal{D}_{xv}^{\text{remain}} \neq \emptyset$ and $k < k_{\max}$}
	        \State Pick a pair $(x,v)\in \mathcal{D}_{xv}^{\text{remain}}$;
	        \If{$(x,\pi_0(x,v)) \in \mathcal{C}$}
	            \If{$(f_{\pi_0}(x,v,w),v)\in \mathcal{D}_{xv}^+$ for all $w\in\mathcal{W}$}
	                \State $\mathcal{D}_{xv}^+ \gets \mathcal{D}_{xv}^+ \cup \{(x,v)\}$; $\ \mathcal{D}_{xv}^{\text{remain}} \gets \mathcal{D}_{xv}^{\text{remain}} \setminus \{(x,v)\}$;
	            \ElsIf{$(f_{\pi_0}(x,v,w),v)\in \mathcal{D}_{xv}^-$ for some $w\in\mathcal{W}$}
	                \State $\mathcal{D}_{xv}^- \gets \mathcal{D}_{xv}^- \cup \{(x,v)\}$; $\ \mathcal{D}_{xv}^{\text{remain}} \gets \mathcal{D}_{xv}^{\text{remain}} \setminus \{(x,v)\}$;
	            \EndIf
	        \Else
	            \State $\mathcal{D}_{xv}^- \gets \mathcal{D}_{xv}^- \cup \{(x,v)\}$; $\ \mathcal{D}_{xv}^{\text{remain}} \gets \mathcal{D}_{xv}^{\text{remain}} \setminus \{(x,v)\}$;
	        \EndIf
	        \State $k \gets k+1$;
	    \EndWhile
	    \State Choose $\Pi_{\pi_0}$ as any set satisfying $\Pi_{\pi_0}^0 \subseteq \Pi_{\pi_0} \subseteq \mathcal{D}_{xv}^+$.
	\end{algorithmic}
\end{algorithm}

When \eqref{equ:system} is discrete, Algorithm~1 can be used to compute a set $\Pi_{\pi_0}$ that is safe and returnable. The algorithm relies on an initial set of state-reference pairs, denoted by $\Pi_{\pi_0}^0$, that is safe and positively invariant under $\pi_0$, i.e., $(x,v)\in \Pi_{\pi_0}^0$ implies
\begin{subequations}\label{equ:discrete_initial}
\begin{align}
    & (1)\quad (x,\pi_0(x,v))\in \mathcal{C}, \\
    & (2)\quad (f_{\pi_0}(x,v,w),v)\in \Pi_{\pi_0}^0,\ \forall w\in\mathcal{W}.
\end{align}
\end{subequations}
While there may exist many sets satisfying \eqref{equ:discrete_initial}, one particularly convenient choice is the set of \emph{safe steady-state pairs}. As discussed in Section~\ref{sec:MA}, the nominal closed-loop system under $\pi_0$ is assumed to exhibit stability-type behavior; in many applications under a constant reference $v$, this is characterized by convergence toward a steady state $x_v(v)$ as $t\to\infty$. In the disturbance-free case $\mathcal{W}=\{0\}$, one may take $\Pi_{\pi_0}^0$ as the set of pairs $(x_v(v),v)$ satisfying $\big(x_v(v),\pi_0(x_v(v),v)\big)\in\mathcal{C}$. When $\mathcal{W}\neq \{0\}$, states in a neighborhood of $x_v(v)$ may need to be checked and included to ensure the positive invariance of $\Pi_{\pi_0}^0$. Such a set is easy to compute when the map $x_v(v)$ is known; otherwise, it may be constructed using data from steady-state experiments.

In Algorithm~1, $\mathcal{D}_{xv}^+$ collects state-reference pairs that are certified to be safe and that will enter $\Pi_{\pi_0}^0$ in finite time regardless of the disturbance realization (and then remain in $\Pi_{\pi_0}^0$ by the positive invariance of $\Pi_{\pi_0}^0$). The set $\mathcal{D}_{xv}^-$ collects pairs that are certified to be unsafe in the sense that, under some disturbance realization, the closed-loop evolution under $\pi_0$ may lead to a violation of $(x,u)\in\mathcal{C}$. The set $\mathcal{D}_{xv}^{\text{remain}}$ contains the pairs that have not yet been classified.
In particular, if a pair $(x,v)$ is safe (Step~4) and, for all $w\in\mathcal{W}$, its successor pair $(f_{\pi_0}(x,v,w),v)$ lies in $\mathcal{D}_{xv}^+$ (Step~5), then $(x,v)$ is added to $\mathcal{D}_{xv}^+$. If a pair $(x,v)$ is unsafe, or if for some $w\in\mathcal{W}$ its successor lies in $\mathcal{D}_{xv}^-$ (Step~7), then $(x,v)$ is added to $\mathcal{D}_{xv}^-$. Once classified, $(x,v)$ is removed from $\mathcal{D}_{xv}^{\text{remain}}$. The algorithm terminates either when $\mathcal{D}_{xv}^{\text{remain}}$ becomes empty or when the iteration limit $k_{\max}$ is reached. Finally, $\Pi_{\pi_0}$ can be chosen as any set satisfying $\Pi_{\pi_0}^0 \subseteq \Pi_{\pi_0} \subseteq \mathcal{D}_{xv}^+$. Since every pair in $\mathcal{D}_{xv}^+$ is guaranteed to safely and eventually enter $\Pi_{\pi_0}^0$, any such choice of $\Pi_{\pi_0}$ is safe and returnable. While $\Pi_{\pi_0}=\mathcal{D}_{xv}^+$ is a trivial valid choice, other choices may be preferred, e.g., for simpler storage. We also note that although $\Pi_{\pi_0}^0$ is already safe and returnable (as it is safe and positively invariant), it may be conservative and small (e.g., when it consists only of safe steady-state pairs). Algorithm~1 enlarges the set and thereby provides the AG algorithm \eqref{equ:AG_1}--\eqref{equ:AG_3} with greater flexibility in selecting safe actions.

\section{Safe Online Learning}\label{sec:SOL}

\begin{figure}
	\centering
		\includegraphics[width=\columnwidth]{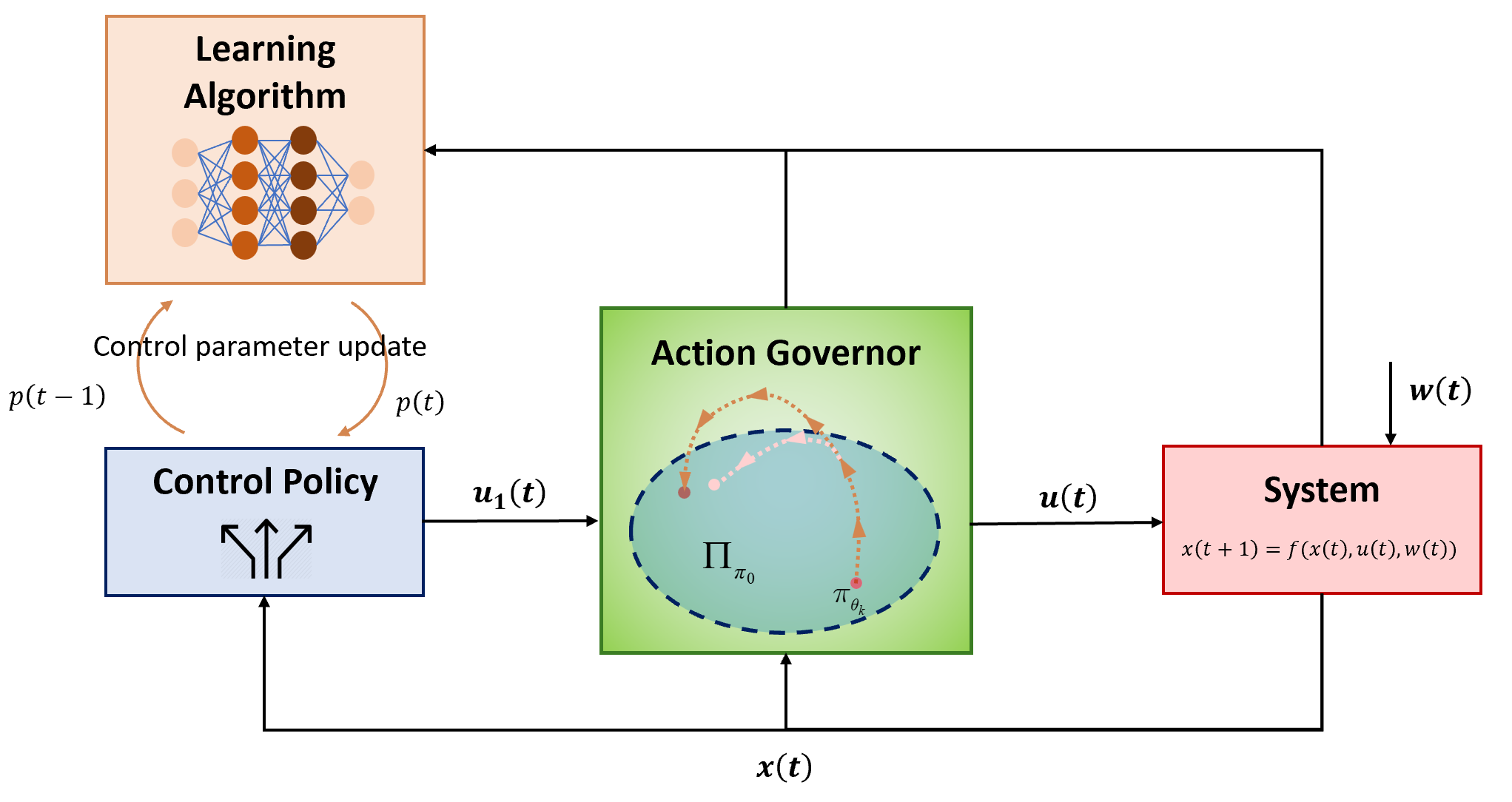}
	\caption{Safe online learning architecture.}
	\label{fig:1}
\end{figure}

The AG can be used to enable safe online learning, i.e., learning while satisfying prescribed safety specifications such as \eqref{equ:constraint}. Fig.~\ref{fig:1} illustrates the proposed safe learning architecture. A learning algorithm updates (or adapts) the control policy online to improve performance and/or to cope with changes in system parameters and operating conditions. Such learning procedures typically rely on injecting {\it excitation signals} (or {\it exploratory actions}) to acquire informative data; however, these exploratory actions may drive the system to violate the safety specifications, even when the baseline policy is designed to satisfy them under nominal operation. To prevent safety violations during learning, the AG serves as a safety supervisor placed between the policy and the plant: it monitors the action proposed by the learning policy and modifies potentially unsafe actions into safe ones.

Under this architecture, various learning algorithms can be combined with the AG to achieve safe online learning. In what follows, we present two representative examples based on different learning algorithms.

\subsection{Safe $Q$-learning}\label{sec:SOL_Q}
$Q$-learning is a classical RL algorithm \cite{watkins1992q} and serves as the foundation of many more advanced RL algorithms \cite{riedmiller2005neural,hasselt2010double}. Given a reward function $r: \mathcal{X} \times \mathcal{U} \to \mathbb{R}$ that measures the immediate performance of each state-action pair $(x,u)$, the $Q$-learning algorithm aims to estimate the optimal $Q$-value of each pair $(x,u)$, which is defined as
\begin{equation}\label{equ:Q_learning_1}
    Q^*(x,u) = r(x,u) + \gamma\, \mathbb{E}\big\{V^*(x') \,\big|\, x, u \big\}
\end{equation}
where $\gamma \in (0,1)$ is a discount factor to ensure boundedness of the $Q$-value, the expectation $\mathbb{E}\{\cdot\}$ is taken over the next state $x'$ conditioned on the current state-action pair $(x,u)$, and the optimal $V$-value of each state is defined as
\begin{equation}\label{equ:Q_learning_2}
    V^*(x) = \max_{\pi}\, \mathbb{E}\left\{\sum_{t = 0}^{\infty} \gamma^t r\big(x(t), u(t)\big) \,\Big|\, x(0) = x, u \sim \pi \right\}
\end{equation}
where the expectation $\mathbb{E}\{\cdot\}$ is taken over all trajectories $\{x(0), u(0), x(1), u(1), ...\}$ conditioned on the given initial state $x(0) = x$ and control policy $\pi$ (i.e., the control action $u(t)$ is selected according to $\pi$ at each time $t$). After a sufficiently accurate estimate of the optimal $Q$-value for each state-action pair $(x,u)$ is obtained, an optimal control policy is calculated according to
\begin{equation}\label{equ:Q_learning_3}
    \pi^*(x) \in \text{argmax}_{u} Q^*(x,u).
\end{equation}
In other words, the ultimate goal of $Q$-learning is to obtain a control policy $\pi^*$ that maximizes the expectation of infinite-horizon discounted cumulative reward $\sum_{t = 0}^{\infty} \gamma^t r(x(t), u(t))$.

In order for the estimated $Q$-values to converge to the optimal $Q$-values, the algorithm must be able to explore different actions at each given state \cite{watkins1992q}. A classical exploration strategy is called {\it $\varepsilon$-greedy} \cite{whitehead1991learning}, where
\begin{equation}\label{equ:Q_learning_4}
    u(t) \in \begin{cases}\, \text{argmax}_{u} \tilde{Q}(x(t),u) & \text{with prob.} = 1 - \varepsilon \\ \,\text{random action in }\,\mathcal{U} & \text{with prob.} = \varepsilon \end{cases}
\end{equation}
i.e., the algorithm has a large probability of $1 - \varepsilon$ to take an action that is optimal according to the latest estimates of the $Q$-values at the current state $x(t)$, and it has a small probability of $\varepsilon$ to take an arbitrary action in the action space $\mathcal{U}$.

During the learning process, violations of prescribed safety conditions may occur, due to, e.g., the application of random exploratory actions. A common strategy is to impose a penalty for such violations so that the control policy gradually learns to satisfy the safety conditions. However, such a strategy learns from safety violations and thus cannot avoid the occurrence of safety violations during learning. In addition, even after a sufficient learning phase and no longer applying any further exploratory actions, safety violations may still occur, due to, e.g., errors in the estimated $Q$-values or the fact that maximizing the expectation of reward allows small (but non-zero) probability of safety violations and penalties. An integration of $Q$-learning and an AG according to Fig.~\ref{fig:1} can avoid the application of any unsafe actions while maintaining the ability of $Q$-learning to evolve the control policy to improve performance. Such an integration is presented in Algorithm~2.

\begin{algorithm}
	\caption{Safe $Q$-learning using the action governor}
	\begin{algorithmic}[1]
	    \State Initialize $Q$-function estimate, $\tilde{Q}(x,u)$; create empty buffer, $\mathcal{D}$; initialize state, $x(0)$; 
	    \For{$T = 0, 1, ..., T_{\max} - 1$}
	    \For{$t = T t_{\text{batch}}, ..., (T+1) t_{\text{batch}} - 1$}
	    \State Select an action $u_1(t)$ according to \eqref{equ:Q_learning_4};
	    \State Adjust $u_1(t)$ to a safe action $u(t)$ using \eqref{equ:AG_1}-\eqref{equ:AG_3};
	    \State Apply $u(t)$ to system and observe next state $x(t+1)$;
	    \State Evaluate a modified immediate reward according to
	    $$
	    \quad \tilde{r}(x(t), u_1(t)) = r(x(t), u(t)) - M \text{\sf dist}_{x(t)}(u_1(t), u(t))
	    $$
	    where $M > 0$ is a large penalty coefficient;
	    \State Calculate a new estimate of the $Q$-value of state-action pair $(x(t), u_1(t))$ according to
	    \begin{align*}
	    Q(x(t), u_1(t)) &= (1 - \alpha) \tilde{Q}(x(t), u_1(t)) \\
	    &\!\!\!\!\!\! + \alpha \big(\tilde{r}(x(t), u_1(t)) + \gamma \max_{u \in \mathcal{U}} \tilde{Q}(x(t+1), u) \big)
	    \end{align*}
	    where $\alpha \in (0,1]$ is a learning rate;
	    \State Store $Q(x(t), u_1(t))$ in buffer $\mathcal{D}$;
	    \EndFor
	    \State Update the $Q$-function estimate $\tilde{Q}(x,u)$ according to data in buffer $\mathcal{D}$;
	    \State Empty the buffer $\mathcal{D}$;
		\EndFor 
	\end{algorithmic} 
\end{algorithm} 

Algorithm~2 updates a $Q$-function estimate, $\tilde{Q}(x,u)$, in a mini-batch manner, where $T_{\max} > 0$ represents the maximum number of batch updates and $t_{\text{batch}} > 0$ represents the batch size. In Step~7, we consider a modified reward function $\tilde{r}(x,u_1)$, which is defined as the original reward $r(x,u)$ minus a penalty for the difference between the original action, $u_1$, and the action after AG adjustment, $u$. This modified reward function achieves two goals: 1) If the original action $u_1$ is safe and the AG passes $u_1$ through (i.e., $u = u_1$), then the modified reward $\tilde{r}(x,u_1)$ is equal to the original reward $r(x,u)$. 2) If the original action $u_1$ is unsafe and the AG adjusts $u_1$ to another action $u$, then the algorithm is informed by the penalty and will learn not to choose the unsafe action $u_1$ at the state $x$.

\subsection{Safe data-driven Koopman control}\label{sec:SOL_K}

Koopman operator theory provides a framework for representing nonlinear dynamics using a (typically higher-dimensional) linear model, enabling the use of linear control techniques such as LQR and linear MPC~\cite{korda2018linear}. For clarity of exposition, consider first a disturbance-free system
\begin{equation*}
    x(t+1)=f(x(t),u(t)),
\end{equation*}
and a collection of functions $g=\{g_1,g_2,\ldots\}$, where each $g_i:\mathcal{X}\to\mathbb{R}$ is an \emph{observable}. The goal is to identify matrices $(A,B)$ such that
\begin{equation}\label{equ:koopman_1}
    g(x(t+1)) \approx A\,g(x(t)) + B\,u(t).
\end{equation}
Defining the lifted state $z=g(x)$ yields the linear predictor
\begin{equation}\label{equ:koopman_2}
    z(t+1)=A z(t)+B u(t),
\end{equation}
which can be used for control design. In practice, the original state $x$ is typically included among the observables to facilitate interpretation and control synthesis~\cite{korda2018linear}. In the presence of uncertainty/disturbance as in \eqref{equ:system}, the mismatch induced by $w(t)$ can be viewed as part of the modeling error handled by the safety supervisor introduced below.

While the computational appeal of Koopman-based control is well recognized, several challenges remain for practical deployment. A key challenge is \emph{guaranteed safety}, i.e., strict satisfaction of prescribed state and control constraints. Even when constraints are enforced in the control synthesis based on a learned Koopman linear model, constraint violations may still occur when the computed control is applied to the original nonlinear system, due to model mismatch. Such mismatch typically arises from: (i) the use of a finite-dimensional approximation of an (infinite-dimensional) Koopman operator, which introduces approximation residuals; and (ii) data-driven identification, where the model is fit on sampled trajectories and the error at unsampled regions is generally unknown \cite{korda2018linear,bruder2020data}. Integrating the Koopman method with an AG, as in Fig.~\ref{fig:1}, provides a mechanism to enforce safety despite model mismatch.

The AG algorithm \eqref{equ:AG_1}--\eqref{equ:AG_3} can be viewed as a (generally nonlinear) map $\pi_a$ that takes the current state $x(t)$ and a proposed action $u_1(t)$ as inputs and outputs the applied action $u(t)$, i.e.,
\begin{equation}\label{equ:koopman_3}
   u(t)=\pi_a\big(x(t),u_1(t)\big).
\end{equation}
With the AG in the loop, the closed-loop dynamics can be written as
\begin{align}\label{equ:koopman_4}
   x(t+1) &= f_a\big(x(t),u_1(t),w(t)\big) \nonumber\\
          &= f\big(x(t),\pi_a\big(x(t),u_1(t)\big),w(t)\big).
\end{align}
Note that even if $f$ is linear, the composite dynamics $f_a$ are typically nonlinear due to the nonlinear action adjustment $\pi_a$. One may therefore apply the Koopman method to approximate $f_a$ by a linear model in a lifted space.

A common approach to estimating a Koopman linear model is data-driven. Suppose a set of observables $g$ has been selected and trajectory data $\{x^k(t),u_1^k(t),x^k(t+1)\}_{k=1}^{k_{\max}}$ have been collected. One can estimate $(A,B)$ by fitting \eqref{equ:koopman_2} in the least-squares sense:
\begin{equation}\label{equ:koopman_5}
   \min_{A,B}\ \big\|Z^+ - A Z - B U_1 \big\|_F,
\end{equation}
where $Z^+=[g(x^1(t+1)),\ldots,g(x^{k_{\max}}(t+1))]$, $Z=[g(x^1(t)),\ldots,g(x^{k_{\max}}(t))]$, $U_1=[u_1^1(t),\ldots,u_1^{k_{\max}}(t)]$, and $\|\cdot\|_F$ denotes the Frobenius norm. An analytical solution is
\begin{equation}\label{equ:koopman_6}
   [A,\ B] = Z^+ \begin{bmatrix} Z \\ U_1 \end{bmatrix}^{\dagger},
\end{equation}
where $(\cdot)^{\dagger}$ denotes the Moore--Penrose inverse.

The estimate $(A,B)$ can also be updated online in a recursive manner~\cite{calderon2021koopman}. Let $(A_{t-1},B_{t-1})$ denote the estimate at time $t-1$. When a new data point $(x(t-1),u_1(t-1),x(t))$ becomes available at time $t$, the estimate is updated as
\begin{equation}\label{equ:koopman_7}
    [A_t,B_t] = [A_{t-1},B_{t-1}] + \varepsilon(t)\,\gamma(t),
\end{equation}
where $\varepsilon(t)=g(x(t)) - A_{t-1}g(x(t-1)) - B_{t-1}u_1(t-1)$ is the prediction error and $\gamma(t)$ is a correction vector computed via
\begin{align}
    \gamma(t) &= \frac{\begin{bmatrix} g(x(t-1)) \\ u_1(t-1) \end{bmatrix}^{\top} \Gamma(t)}{\begin{bmatrix} g(x(t-1)) \\ u_1(t-1) \end{bmatrix}^{\top} \Gamma(t)\begin{bmatrix} g(x(t-1)) \\ u_1(t-1) \end{bmatrix} + \lambda}, \label{equ:koopman_8}\\[4pt]
    \Gamma(t+1) &= \frac{1}{\lambda}\,\Gamma(t)\left(I - \begin{bmatrix} g(x(t-1)) \\ u_1(t-1) \end{bmatrix}\gamma(t)\right), \label{equ:koopman_9}
\end{align}
where $\lambda\in(0,1]$ is a forgetting factor ($\lambda=1$ corresponds to no forgetting). This recursive procedure enables online adaptation of the Koopman model using operating data, so that the model tracks changes in system parameters and environmental conditions.An integration of online learning Koopman model, model-based determination of control, and an AG to enforce safety is presented in Algorithm~3.

\begin{algorithm}
	\caption{Safe data-driven Koopman control using the action governor}
	\begin{algorithmic}[1]
	    \State Select observables $g$; initialize Koopman model $(A_0,B_0)$; initialize $x(0)$ and $z(0)=g(x(0))$;
	    \For{$t = 0,1,2,\ldots$}
	        \State Compute a proposed action $u_1(t)$ using the linear predictor $z(t+k+1)=A_t z(t+k)+B_t u_1(t+k)$, $k=0,1,\ldots$ (e.g., via \eqref{equ:koopman_10});
	        \State Adjust $u_1(t)$ to an applied safe action $u(t)$ using \eqref{equ:AG_1}--\eqref{equ:AG_3};
	        \State Apply $u(t)$ to the system and observe $x(t+1)$;
	        \State Compute $z(t+1)=g(x(t+1))$;
	        \State Update $(A_t,B_t)$ to $(A_{t+1},B_{t+1})$ using \eqref{equ:koopman_7}--\eqref{equ:koopman_9};
	    \EndFor
	\end{algorithmic}
\end{algorithm}

In Step~3, given the linear predictor $z(t+k+1)=A_t z(t+k)+B_t u_1(t+k)$, various strategies can be used to compute the proposed action $u_1(t)$. Here we consider a stabilization setting and use an unconstrained finite-horizon LQR/MPC formulation. At each time step $t$, we compute a sequence $\{u_1(k|t)\}_{k=0}^{N-1}$ by solving

\begin{subequations}\label{equ:koopman_10}
\begin{align}
    \min_{u_1(\cdot|t)} \sum_{k=0}^{N-1}\Big(z(k|t)^{\top}Q z(k|t) &+ u_1(k|t)^{\top}R u_1(k|t)\Big)  \nonumber \\ 
    + &z(N|t)^{\top}Q_f z(N|t)  \\
    \text{s.t. } z(k+1|t)=A_t &z(k|t)+B_t u_1(k|t),\label{equ:koopman_10b}\\
    z(0|t)&=g(x(t)), \label{equ:koopman_10c}
\end{align}
\end{subequations}
where $Q$, $R$, and $Q_f$ are weighting matrices and $N$ is the prediction horizon. The applied proposal is $u_1(t)=u_1(0|t)$, and the procedure is repeated in a receding-horizon manner. This choice is motivated by two considerations. First, since \eqref{equ:koopman_10} contains only equality constraints, it admits an efficient solution (indeed, it is a discrete-time finite-horizon LQR problem)~\cite{lewis2012optimal}. Second, we use a finite-horizon formulation rather than the infinite-horizon LQR because convergence of the infinite-horizon solution requires controllability-type conditions on $(A_t,B_t)$~\cite{lewis2012optimal}, which may not always hold during online data-driven updates.

Finally, safety of the \emph{applied} action is enforced by the AG. In particular, under the feasibility conditions of Proposition~1 (e.g., feasibility of \eqref{equ:AG_2} at the initial time), the AG guarantees satisfaction of the constraints \eqref{equ:constraint} for all times, even when the Koopman model is imperfect and the proposed action $u_1(t)$ is computed without explicitly incorporating constraints in \eqref{equ:koopman_10}.

\section{Illustrative Example}\label{sec:Ex}

Consider the following discrete-time system:
\begin{align}\label{equ:ex_1}
    x(t+1) &= A x(t) + B u(t) + E w(t) \\
    &\!\!\!\!\!\!\!\! = \begin{bmatrix} 1 & 1 \\ 0 & 1 \end{bmatrix} x(t) + \begin{bmatrix} 0 \\ 1 \end{bmatrix} u(t) + \begin{bmatrix} 0 \\ 1 \end{bmatrix} w(t), \nonumber 
\end{align}
where $x(t)=[x_1(t),x_2(t)]^\top\in\mathbb{R}^2$ is the state, $u(t)\in\mathbb{R}$ is the control input, and the term $w(t)\in\mathbb{R}$ depends on the state according to
\begin{equation}\label{equ:ex_2}
    w(t)=\sin\big(10x_1(t)\big).
\end{equation}
For a given $w(t)$, \eqref{equ:ex_1} is linear in $(x,u)$; however, due to the state-dependent relation \eqref{equ:ex_2}, the overall system is nonlinear. In particular, the linearization of the closed-loop state-update map $x(t+1)=A x(t)+B u(t)+E\sin(10x_1(t))$ about the origin has Jacobian
\begin{equation*}
    f_x(0)=\begin{bmatrix}1 & 1\\ 10 & 1\end{bmatrix},
\end{equation*}
which differs from the nominal matrix $A$ in \eqref{equ:ex_1}. Moreover, for large $|x_1|$, the term $\sin(10x_1)$ varies rapidly with $x_1$ while remaining bounded in $[-1,1]$.

A practical strategy is to treat \eqref{equ:ex_2} as an unknown but bounded disturbance, i.e., $w(t)\in\mathcal{W}=[-1,1]$, so that linear robust-control and constraint-handling tools can be applied. Following this strategy, one may design a linear stabilizing controller
\begin{equation}\label{equ:ex_3}
    u(t)=Kx(t)=[-0.2054,\,-0.7835]\,x(t),
\end{equation}
where $K$ is the infinite-horizon LQR gain corresponding to $Q=\mathrm{diag}(1,1)$ and $R=10$. However, under the state-dependent disturbance \eqref{equ:ex_2}, this linear controller may only regulate the state to a neighborhood of the origin, as illustrated by the blue trajectory in Fig.~\ref{fig:2}.

We further impose the following state and input constraints:
\begin{equation}\label{equ:ex_4}
    -20 \le x_1(t) \le 20,\quad -4 \le x_2(t) \le 10,\quad -6 \le u(t) \le 6,\quad \forall t.
\end{equation}
The linear controller \eqref{equ:ex_3} does not explicitly handle these constraints. In what follows, we employ the generalized AG to enforce \eqref{equ:ex_4} (under the feasibility conditions established earlier) and integrate it with safe online learning to improve closed-loop performance.

\subsection{Action governor design and safe control}

The first step in designing a generalized AG for enforcing the constraints in \eqref{equ:ex_4} is to specify a nominal policy $\pi_0$ for which a nonempty safe and returnable set exists. We consider the following linear policy:
\begin{equation}\label{equ:ex_5}
    u_0(t)=\pi_0\big(x(t),v(t)\big)=Kx(t)+Lv(t),
\end{equation}
where $K=[-0.2054,-0.7835]$ is the same LQR gain as in \eqref{equ:ex_3} so that the closed-loop matrix $\tilde{A}=A+BK$ is Schur, and
$L=-K\begin{bmatrix}1\\0\end{bmatrix}=0.2054$ is chosen so that, in the disturbance-free case ($w\equiv 0$), the steady state associated with a constant reference $v$ satisfies $x_v(v)=[v,0]^\top$ and $u_0=0$.

\begin{figure}
	\centering
		\includegraphics[width=\columnwidth]{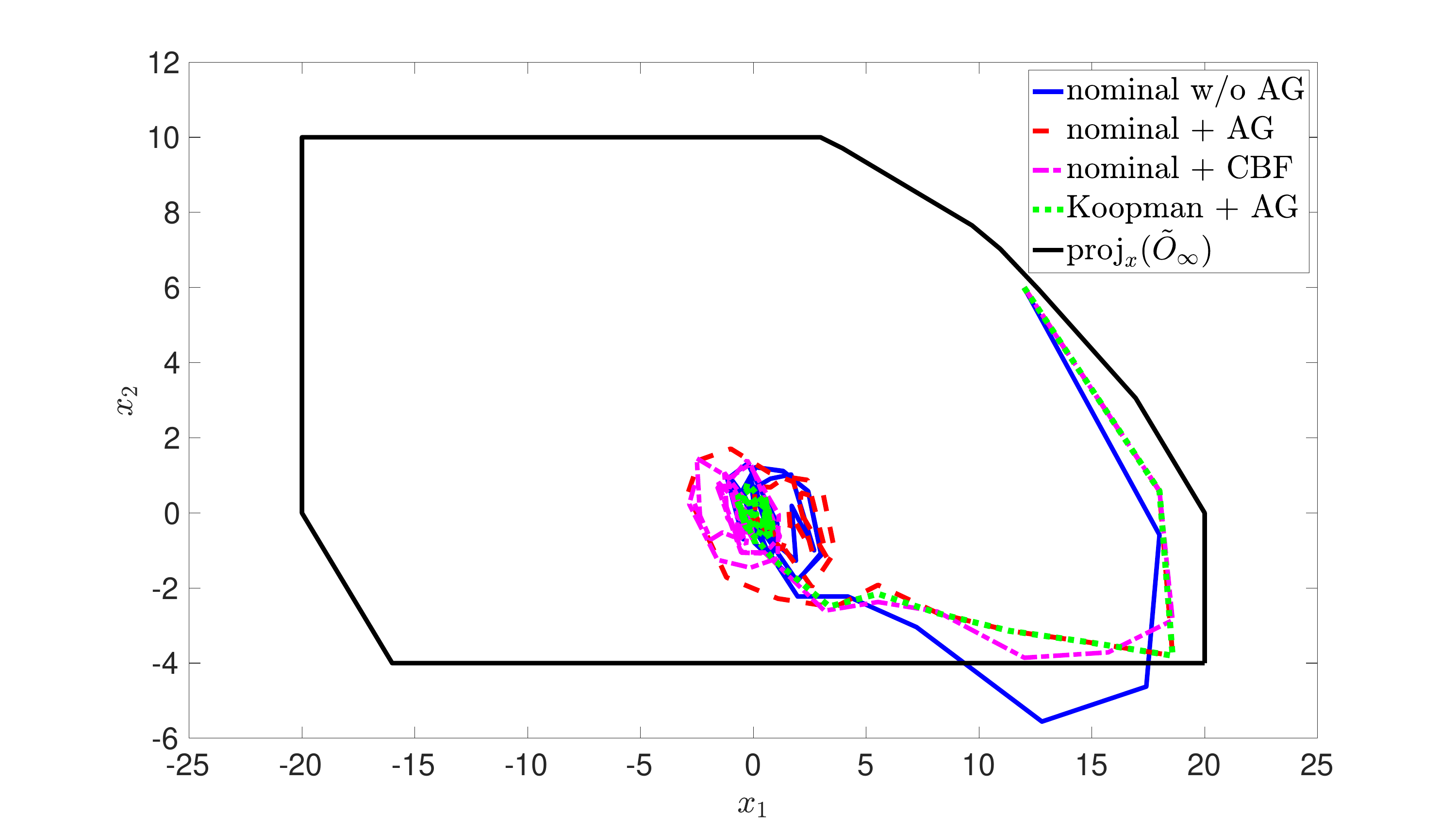}
	\caption{State trajectories under nominal control without AG, nominal control with AG, nominal control with CBF, and learned Koopman control with AG.}
    \label{fig:2}
\end{figure}

Treating \eqref{equ:ex_1} as a linear system with a bounded disturbance $w(t)\in\mathcal{W}=[-1,1]$, and following the linear-systems design in Section~\ref{sec:GAG_linear}, we compute the MOAS $\tilde{\mathcal{O}}_{\infty}$ and set $\Pi_{\pi_0}=\tilde{\mathcal{O}}_{\infty}$. The projection $\text{\sf proj}_x(\tilde{\mathcal{O}}_{\infty})$ is shown in Fig.~\ref{fig:2}, where the black curve indicates its boundary. We then construct an AG using $\Pi_{\pi_0}=\tilde{\mathcal{O}}_{\infty}$ and the online optimization \eqref{equ:AG_4} with $\text{\sf dist}_{x(t)}(u_1(t),u)=|u_1(t)-u|$, so that control actions are adjusted to enforce \eqref{equ:ex_4}.

Fig.~\ref{fig:2} compares the trajectories of \eqref{equ:ex_1} from the initial condition $x(0)=(14,6)$ under the nominal controller \eqref{equ:ex_3} without and with AG supervision (blue and red curves, respectively). Without AG supervision, the trajectory violates the constraints, whereas with AG supervision the applied input satisfies \eqref{equ:ex_4} for all times under the feasibility conditions stated earlier (cf.\ Proposition~1).

For comparison, we also implement a control barrier function (CBF) for this example. Since the system is discrete-time and has relative degree $2$, we adopt the method in \cite{xiong2022discrete} to construct a discrete-time exponential CBF. The resulting trajectory is shown by the magenta curve in Fig.~\ref{fig:2}. Similar to AG supervision, the CBF-guarded trajectory satisfies the constraints in this example. We emphasize, however, that our generalized AG design provides a recursive feasibility guarantee in the presence of simultaneous state and input constraints (Proposition~3), whereas for discrete-time CBF-based filters, feasibility may be lost in general when additional input constraints are imposed, and such a recursive feasibility guarantee is not explicitly available in \cite{xiong2022discrete}. This highlights an important practical advantage of the generalized AG for handling simultaneous state and control constraints.

To illustrate Algorithm~1 for computing $\Pi_{\pi_0}$ for discrete systems, we also construct a finite-state abstraction of the (already discrete-time) closed-loop dynamics on a grid. Specifically, we grid the state range $[-25,25]\times[-10,15]$ with resolution $\Delta x_1\times\Delta x_2=0.5\times 0.5$, the reference range $[-25,25]$ with resolution $\Delta v=0.5$, and the disturbance range $[-1,1]$ with resolution $\Delta w=0.1$. These ranges cover the state and input limits in \eqref{equ:ex_4} as well as the disturbance bound $\mathcal{W}=[-1,1]$.

The discrete transition map $f$ assigns each grid point $(x,v,w)$ to the next grid point $x^+$ closest (in Euclidean distance) to $\tilde{A}x + BLv + Ew$. Recall that Algorithm~1 requires an initial set $\Pi_{\pi_0}^0$ satisfying \eqref{equ:discrete_initial}. To construct a valid $\Pi_{\pi_0}^0$, we consider the ellipsoidal set
\begin{equation}\label{equ:ex_6}
    \mathcal{E}(v)=\left\{x\in\mathbb{R}^2:\ (x-x_v(v))^\top P^{-1}(x-x_v(v))\le 1\right\},
\end{equation}
where $x_v(v)=(I_2-\tilde{A})^{-1}BLv=\begin{bmatrix}v\\0\end{bmatrix}$ is the steady state corresponding to constant $v$ and $w\equiv 0$, and $P$ solves the discrete-time Lyapunov equation
\begin{equation}\label{equ:ex_7}
    \frac{1}{\alpha}\tilde{A}P\tilde{A}^\top - P + \frac{1}{1-\alpha}EE^\top = 0,
\end{equation}
with $\alpha=0.75$ chosen such that $\rho(\tilde{A})^2 < \alpha < 1$. By Theorem~3 of \cite{polyak2006rejection}, $\mathcal{E}(v)$ is positively invariant for the dynamics $x^+ = \tilde{A}x + BLv + Ew$ under constant $v$ and $w\in\mathcal{W}=[-1,1]$. Therefore, we define $\Pi_{\pi_0}^0$ as the collection of all grid pairs $(x,v)$ that belong to
$\bigcup_v(\mathcal{E}(v), v)$ and satisfy $(x,\pi_0(x,v))\in\mathcal{C}$. With this $\Pi_{\pi_0}^0$, we then run Algorithm~1 to compute $\Pi_{\pi_0}$.

\begin{figure}
	\centering
		\includegraphics[width=\columnwidth]{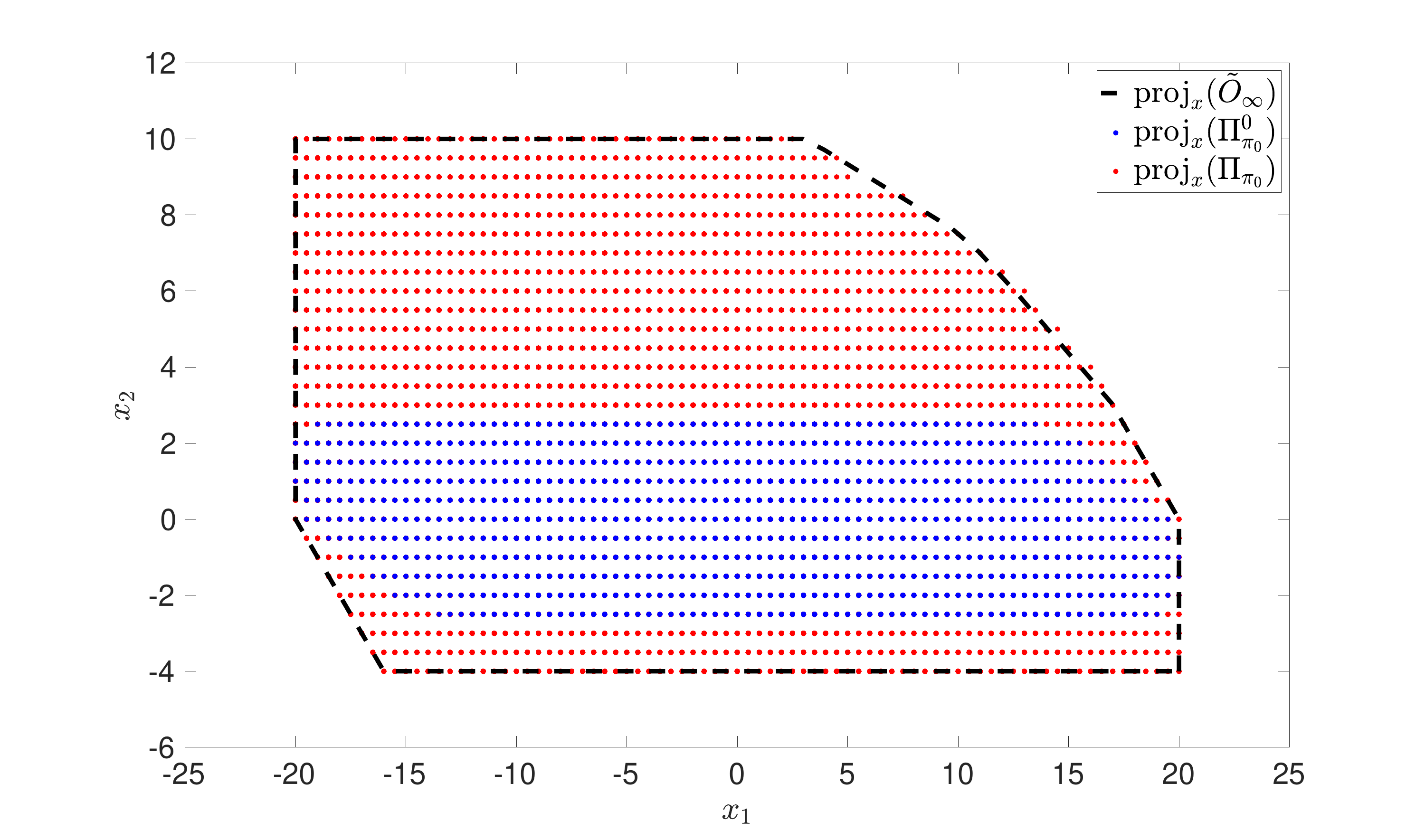}
	\caption{Safe sets computed using linear systems approach versus discrete systems approach.}
    \label{fig:3}
\end{figure}

Fig.~\ref{fig:3} shows the projections of $\Pi_{\pi_0}^0$ and $\Pi_{\pi_0}$ onto the state space as blue and red points, respectively; the boundary of $\text{\sf proj}_x(\tilde{\mathcal{O}}_{\infty})$ computed via the linear-systems approach is shown by the black dashed curve. Two observations can be made. First, Algorithm~1 significantly enlarges the safe set from $\Pi_{\pi_0}^0$ (blue) to $\Pi_{\pi_0}$ (red). Second, the final set computed via the discrete abstraction closely matches that computed via the linear-systems approach; the small discrepancies are mainly due to quantization effects when mapping the next state to the nearest grid point. These results verify the effectiveness of Algorithm~1.

\subsection{Safe online learning using action governor}

While the action governor enforces the constraints in \eqref{equ:ex_4}, the closed-loop performance under the nominal LQR controller \eqref{equ:ex_3} remains unsatisfactory. As shown in Fig.~\ref{fig:2}, both the nominal controller without AG supervision (blue) and the nominal controller with AG supervision (red) yield state trajectories $x(t)=(x_1(t),x_2(t))$ that exhibit noticeable oscillations around the origin, mainly due to the state-dependent nonlinearity in \eqref{equ:ex_2}. In many practical applications, such nonlinearities (and, more generally, state-dependent disturbances) are a priori unknown or difficult to model accurately. In such cases, a viable approach to improving performance is safe online learning. Since safe $Q$-learning (Section~\ref{sec:SOL_Q}) has been demonstrated on multiple occasions, including \cite{li2021robust,li2021safereinforcementlearningusing}, we focus here on illustrating the newly proposed safe data-driven Koopman control method in Section~\ref{sec:SOL_K}.

Inspired by the observable-selection approach based on higher-order derivatives of nonlinear dynamics in~\cite{mamakoukas2021derivative}, for system \eqref{equ:ex_1}--\eqref{equ:ex_2} we choose the following observables:
\begin{equation}\label{equ:ex_8}
    z(t)=g(x(t))=
    \begin{bmatrix}
    x_1(t)\\
    x_2(t)\\
    \sin\!\big(10x_1(t)\big)\\
    \sin\!\big(10x_1(t)+10x_2(t)\big)
    \end{bmatrix}.
\end{equation}
We then implement Algorithm~3 with the initial lifted model
\begin{equation}\label{equ:ex_9}
    A_0=
    \begin{bmatrix}
    A & 0_{2\times 2}\\
    0_{2\times 2} & 0_{2\times 2}
    \end{bmatrix},
    \quad
    B_0=
    \begin{bmatrix}
    B\\
    0_{2\times 1}
    \end{bmatrix},
\end{equation}
so that the initial model corresponds to the nominal linear dynamics and does not account for the nonlinearity. At each time step, we generate the \emph{pre-adjustment} control $u_1(t)$ via an infinite-horizon LQR designed on the lifted linear model, using $Q'=\mathrm{diag}(1,1,0,0)$ and $R'=10$. This choice is made to minimize the effect of using different cost weights when comparing performance, recalling that the nominal controller \eqref{equ:ex_3} is also an infinite-horizon LQR with $Q=\mathrm{diag}(1,1)$ and $R=10$. (Equivalently, the infinite-horizon LQR law can be obtained from the finite-horizon formulation \eqref{equ:koopman_10} by choosing $Q_f$ as the solution to the discrete-time algebraic Riccati equation.) The applied input to the plant is $u(t)$ obtained after AG adjustment, consistent with Algorithm~3.

\begin{figure}
	\centering
		\includegraphics[width=\columnwidth]{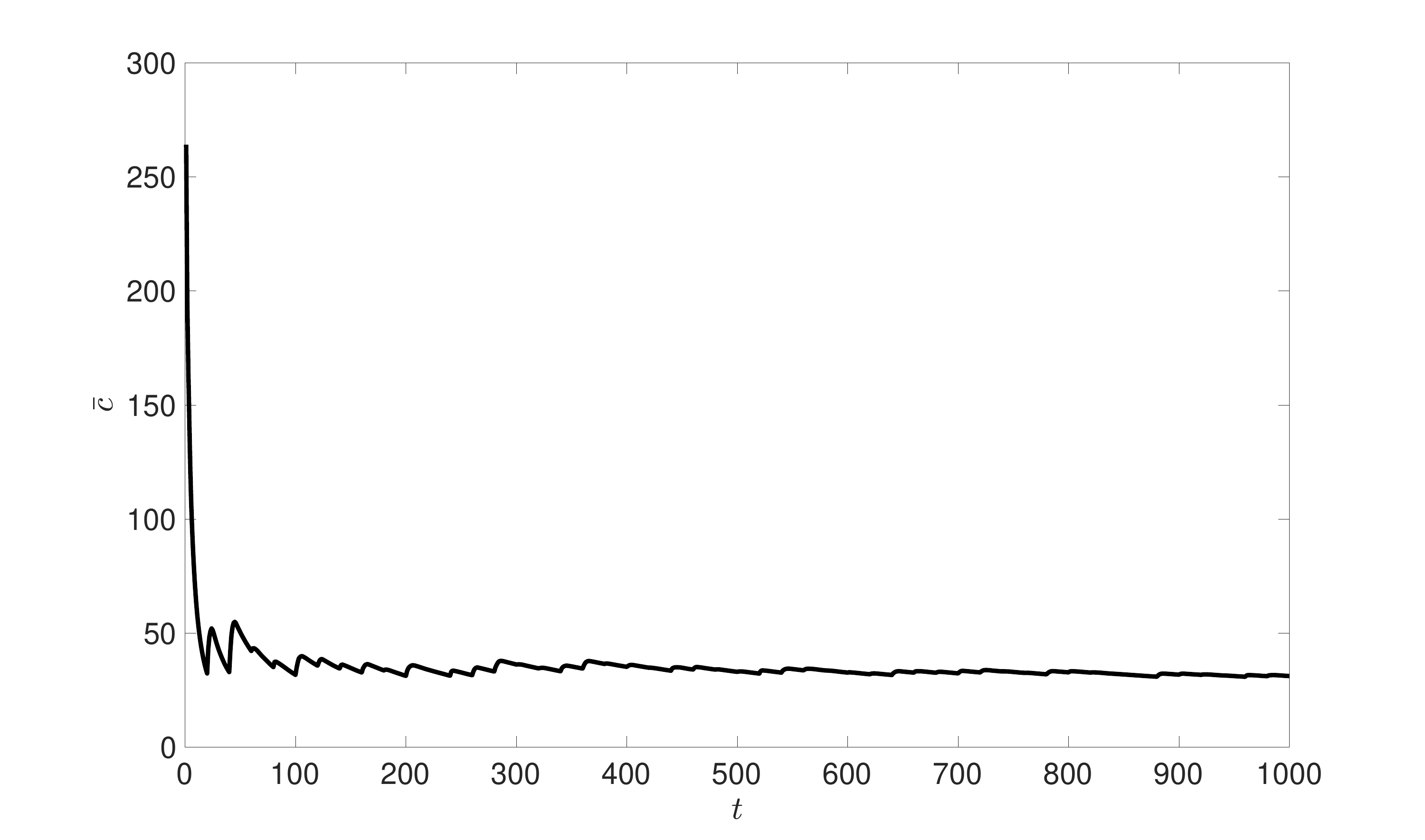}
	\caption{Average cost during learning.}
    \label{fig:4}
\end{figure}

To improve coverage of the safe operating region in simulation, we restart trajectories by reinitializing the state uniformly at random in $\text{\sf proj}_x(\Pi_{\pi_0})$ every $\Delta t=20$ time steps. To monitor performance during online learning, we use the single-step LQR cost
\begin{equation}\label{equ:ex_10}
    c(t)=x(t)^\top Q x(t)+u(t)^\top R u(t)=\|x(t)\|_2^2+10|u(t)|^2,
\end{equation}
and compute the running average
\begin{equation}\label{equ:ex_11}
    \bar{c}(t)=\frac{1}{t+1}\sum_{k=0}^{t} c(k).
\end{equation}
The evolution of $\bar{c}(t)$ is shown in Fig.~\ref{fig:4}. The average cost decreases and converges as learning proceeds, indicating improved control performance. In our simulations, the AG optimization remained feasible (with feasible initializations), and hence no constraint violation was observed throughout learning; therefore, a separate plot of constraint violations is omitted.

Finally, Fig.~\ref{fig:2} compares the trajectories of \eqref{equ:ex_1} from $x(0)=(14,6)$ under the nominal controller \eqref{equ:ex_3} with AG supervision (red) and under the control determined from the learned Koopman model with AG supervision (green). The learned Koopman controller drives and maintains the state within a substantially smaller neighborhood of the origin than the nominal controller, illustrating the effectiveness of safe online learning for improving control performance under the same safety constraints.

\section{Conclusions}\label{sec:conclude}

This paper introduced a safety supervisor, termed the {\it Generalized Action Governor} (AG), which augments a nominal closed-loop system with the capability to enforce prescribed state and input constraints through online action adjustment (under the feasibility conditions established in the paper). We presented the generalized AG for general discrete-time systems and analyzed its key properties, highlighting the use of a safe set with {\it returnability} (rather than requiring positive invariance). Based on this theory, we developed tailored AG design procedures for linear systems and for discrete systems with finite state and action spaces under bounded uncertainties. We further discussed safe online learning enabled by the AG and presented two learning-based controllers---safe $Q$-learning and safe data-driven Koopman control---both integrated with the generalized AG. Numerical results illustrated the proposed methods and suggested that the generalized AG offers a practical and general framework for safe autonomy.

\printcredits

\section*{Declaration of competing interest}
The authors declare that they have no known competing financial interests or personal relationships that could have appeared to influence the work reported in this paper.

\section*{Data availability}
Data will be made available on request.

\bibliographystyle{model1-num-names}

\bibliography{refs}

\section*{Author biography}
\bio{Peiyuan_Fang}
\noindent\textbf{Peiyuan Fang} received his B.E. degree in vehicle engineering in 2019, Tongji University, Shanghai, China. Currently, he is working toward the Ph.D. degree at the Institute of Intelligent Vehicles, Tongji University, Shanghai, China. His research interests in decision-Making, planning, and control for autonomous vehicles in complex environments.
\endbio
\vspace{100pt}

\bio{Weiqi_Zhang}
\noindent\textbf{Weiqi Zhang} received his B.E. degree in vehicle engineering from Tongji University, Shanghai, China, in 2025. Currently, he is working toward the Ph.D. degree at the Institute of Intelligent Vehicles, Tongji University, Shanghai, China. His research interests include autonomous, racing, safe reinforcement learning, and end-to-end autonomous driving.
\endbio
\vspace{10pt}

\bio{Lu_Xiong}
\noindent\textbf{Lu Xiong} received the Ph.D. degree in vehicle engineering from Tongji University, Shanghai, China, in 2005. He is currently the Vice President and a Professor with the School of Automotive Studies, Tongji University. His current research interests include the dynamic control of distributed drive electric vehicles, motion planning and control of intelligent vehicles, and all-terrain vehicles. He won the First Prize in the Shanghai Science and Technology Progress Awards in 2013, 2020, and 2022. He was a recipient of the National Science Fund for Distinguished Young Scholars.
\endbio

\bio{Nan_Li}
\noindent\textbf{Nan Li} received the B.E. degree in vehicle engineering from Tongji University, Shanghai, China, in 2014, the M.S. degree in mechanical engineering, the M.S. degree in mathematics, and the Ph.D. degree in aerospace engineering from the University of Michigan, Ann Arbor, MI, USA, in 2016, 2020, and 2021, respectively. Dr. Li is currently a Professor with the School of Automotive Studies at Tongji University. Prior to joining Tongji in 2024, he was a Postdoc Fellow at the University of Michigan, Ann Arbor, from 2021 to 2022, and a tenure-track Assistant Professor at Auburn University, AL, USA, from 2022 to 2024. His research interests are in safety-critical control, optimal and predictive control, learning, multi-agent systems, and their applications in automotive and aerospace systems.
\endbio

\bio{Yanjun_Huang}
\textbf{Yanjun Huang} is a Professor at School of Automotive studies, Tongji University. He received his PhD Degree in 2016 from the Department of MME at University of Waterloo. His research interest is mainly on autonomous driving and artificial intelligence in terms of decision-making and planning, motion control, human-machine cooperative driving. He has published several books, over 80 papers in journals and conference; He is the recipient of IEEE Vehicular Technology Society 2019 Best Land Transportation Paper Award. He is serving as AE of IEEE/TITS, IET/ITS, SAE/IJCV, Springer Book series of connected and autonomous vehicle, etc.
\endbio

\bio{Yutong_Li}
\textbf{Yutong Li} is an ADAS algorithm and software engineer at Ford Motor Company in Dearborn, Michigan. His research interests are in control theory for safety-critical systems, safe reinforcement learning, and their applications to automotive systems. He received his Ph.D. degree in Automotive Engineering from Tsinghua University, Beijing, China, in 2018. Prior to joining Ford in 2022, Dr. Li was with the University of Michigan, Ann Arbor, as a Postdoctoral Research Fellow.
\endbio

\bio{Ilya_Kolmanovsky}
\textbf{Ilya Kolmanovsky} is a Professor in the Department of Aerospace Engineering at the University of Michigan, Ann Arbor, MI, USA, with research interests in control theory for systems with state and control constraints, and in control applications to aerospace and automotive systems. He received his Ph.D. degree in Aerospace Engineering from the University of Michigan in 1995. Prior to joining the University of Michigan as a faculty in 2010, Dr. Kolmanovsky was with Ford Research and Advanced Engineering in Dearborn, Michigan for close to 15 years. He is a Fellow of IEEE and IFAC, and the Editor-in-Chief of IEEE Transactions on Control Systems Technology.
\endbio

\bio{Anouck_Girard}
\textbf{Anouck Girard} received the Ph.D. degree in Ocean Engineering from the University of California, Berkeley, CA, USA, in 2002. She was with the University of Michigan, Ann Arbor, MI, USA, from 2006 to 2024. She joined Embry-Riddle Aeronautical University in 2025 and is currently the Chair of the Department of Aerospace Engineering. Her current research interests include vehicle dynamics and control systems. She has co-authored the book Fundamentals of Aerospace Navigation and Guidance (Cambridge University Press, 2014). 
\endbio

\bio{Hongtei_Eric_Tseng}
\textbf{Hongtei Eric Tseng} received his B.S. degree from National Taiwan University, Taipei, Taiwan, in 1986. He received his M.S. and Ph.D. degrees from the University of California, Berkeley, in 1991 and 1994, respectively, all in Mechanical Engineering. He was a Senior Technical Leader of Controls and Automated Systems in Research and Advanced Engineering at Ford. Many of his contributed technologies led to production vehicles implementation. His technical achievements have been recognized with the highest technical award internally -- the Henry Ford Technology Award -- seven times, as well as externally by the American Automatic Control Council with Control Engineering Practice Award in 2013. Dr.~Tseng is a member of NAE. He has over 100 US patents, one third of which are in production, and is the author/coauthor of over 120 publications, including 6 book chapters.
\endbio
\vspace{20pt}

\bio{Dimitar_Filev}
\textbf{Dimitar Filev} was Senior Henry Ford Technical Fellow in Control and AI with Research \& Advanced Engineering of Ford Motor Company. His research is in computational intelligence, AI and intelligent control, and their applications to autonomous driving, vehicle systems, and automotive engineering. He holds over 100 granted US patents and has been awarded with the IEEE SMCS 2008 Norbert Wiener Award and the 2015 Computational Intelligence Pioneer's Award. Dr.~Filev is a Fellow of IEEE and a member of NAE. He was President of the IEEE Systems, Man, \& Cybernetics Society.
\endbio

\end{document}